\documentclass[a4paper,fleqn,usenatbib]{mnras}

\usepackage{newtxtext,newtxmath}

\usepackage[T1]{fontenc}
\usepackage{ae,aecompl}

\usepackage{subcaption}
\usepackage{graphicx, nicefrac}	
\usepackage{multicol}
\usepackage{amsmath}	
\usepackage{nicefrac}
\usepackage{tablefootnote}
\usepackage{threeparttable}
\usepackage{nccmath}
\usepackage{xcolor}
\usepackage{upgreek}
\usepackage{dcolumn}
\usepackage{hyperref}
\usepackage{multirow}
\usepackage{booktabs}
\usepackage{braket}
\usepackage{siunitx}

\title[Warped disc of DoAr\,44]{Radio-continuum decrements associated to shadowing from the central warp in transition disc DoAr\,44}

\author[C. Arce-Tord et al.]{Carla Arce-Tord,\!$^{1,2,5}$\thanks{E-mail: carla.atastro@gmail.com} Simon Casassus,\!$^{1,5,14}$ William R. F. Dent,\!$^{3}$ Sebastián Pérez,\!$^{4,5,8}$ \newauthor Miguel Cárcamo,\!$^{6,7,8}$ Philipp Weber,\!$^{4,5,8}$ Natalia Engler,\!$^{9}$ Lucas A. Cieza,\!$^{5,10}$ \newauthor Antonio Hales,\!$^{3,12}$ Alice Zurlo,\!$^{5,10,11}$  Sebastian Marino \!$^{13}$ \\\\
$^{1}$ Departamento de Astronom\'{\i}a, Universidad de Chile, Casilla 36-D, Santiago, Chile\\
 $^{2}$European Southern Observatory, Alonso de Cordova 3107, Casilla 19, Santiago, Chile \\
 $^{3}$Joint ALMA Observatory, Alonso de Cordova 3107, Casilla 19, Santiago, Chile.\\
 $^{4}$Departamento de Física, Universidad de Santiago de Chile, Av. Victor Jara 3659, Santiago\\
 $^{5}$Millennium Nucleus on Young Exoplanets and their Moons (YEMS), Chile.\\
$^{6}$Jodrell Bank Centre for Astrophysics, Department of Physics and Astronomy, University of Manchester, Manchester, UK \\
$^{7}$University of Santiago of Chile (USACH), Faculty of Engineering, Computer Engineering Department, Chile \\
$^{8}$Center for Interdisciplinary Research in Astrophysics and Space Exploration (CIRAS), Universidad de Santiago de Chile\\
$^{9}$ETH Zurich, Institute for Particle Physics and Astrophysics, Wolfgang-Pauli-Strasse 27, CH-8093 Zurich, Switzerland \\
$^{10}$N\'ucleo de Astronom\'ia, Facultad de Ingenier\'ia y Ciencias, Universidad Diego Portales, Av. Ejercito 441, Santiago, Chile\\
$^{11}$Escuela de Ingenier\'ia Industrial, Facultad de Ingenier\'ia y Ciencias, Universidad Diego Portales, Av. Ejercito 441, Santiago, Chile \\
 $^{12}$National Radio Astronomy Observatory, 520 Edgemont Road, Charlottesville, VA 22903-2475, United States of America.\\
$^{13}$Department of Physics and Astronomy, University of Exeter, Stocker Road, Exeter, EX4 4QL, UK. \\
$^{14}$ Data Observatory Foundation, Chile
}

\date{Accepted XXX. Received YYY; in original form ZZZ}

\pubyear{2023}

\begin{document}

\label{firstpage}
\pagerange{\pageref{firstpage}--\pageref{lastpage}}
\maketitle

\label{firstpage}

\begin{abstract}
Warps have often been used to explain disc properties, but well characterised examples are important due to their role in disc evolution. Scattered light images of discs with central gaps have revealed sharp warps, such that the outer rings are shadowed by tilted inner discs. The near-IR intensity drops along the ring around TTauri star DoAr\,44 have been interpreted in terms of a central warp. We report new ALMA observations of DoAr\,44 in the continuum at 230\,GHz and 350\,GHz (at $\sim$10\,au), along with a new epoch of SPHERE/IRDIS differential polarised imaging taken during excellent weather conditions. The ALMA observations resolve the ring and confirm the decrements proposed from deconvolution of coarse 336\,GHz data. The scattered light image constrains the dips, which correspond to a misaligned inner disc with a relative inclination $\xi$\,=\,21.4\,$^{+6.7}_{-8.3}$\,deg. The SPHERE intensity profile shows a morphological change compared to a previous epoch that may be interpreted as a variable orientation of the inner disc, from $\xi$\,$\sim$\,30\,deg to $\xi$\,$\sim$\,20\,deg. The intensity dips probably correspond to temperature decrements, as their mm-spectral index, $\alpha^{230 \textrm{GHz}}_{350 \textrm{GHz}} \sim$\,2.0\,$\pm$\,0.1, is indicative of optically thick emission. The azimuth of the two temperature decrements are leading clockwise relative to the IR-dips, by $\eta$\,=\,14.95\,deg and $\eta$\,=\,7.92\,deg. For a retrograde disc, such shifts are expected from a thermal lag and imply gas surface densities of $\Sigma_g$\,=\,117\,$\pm$\,10 g/cm$^2$ and $\Sigma_g$\,=\,48\,$\pm$\,10 g/cm$^2$. A lopsided disc, with contrast ratio $f_r$=2.4\,$\pm$\,0.5, is also consistent with the large continuum crescent.

\end{abstract}


\section{Introduction}

Transition discs \citep[TDs,][]{Espaillat2010} are useful sources for the study of circumstellar disc warps. This is due to their characteristic inner dust cavity, which may have been cleared by the orbit of young planets \citep[e.g.][]{dong15}. Resolved observations in scattered light have revealed azimuthal dips along the outer rings. For instance, infrared (IR) and optical scattered light observations in HD\,142527 \citep{fukagawa+2006, casassus12, avenhaus14} revealed dips in the outer disc thought to be shadows due to a central warp \citep{marino15}. The shadows in this disc, seen in polarised-differential imaging (PDI), are explained by a tilted inner disc, 70$^\circ$ relative to the outer disc. 

Analysis of gas kinematics traced by ALMA observations of HD\,142527 in CO(6$-$5) showed that the intra-cavity gas is close to free-fall through the central warp \citep{casassus15}. This warped structure is most likely caused by the inclined orbit of the low mass companion HD\,142527B \citep{biller12, price18}, which translates into shadows deep enough to cool the dust in the outer ring and cause the local decrements in the ALMA continuum \citep{casassus15b}. 

Observations with the Spectro Polarimetric High-contrast Exoplanet REsearch \citep[SPHERE;][]{beuzit19} instrument at the Very Large Telescope (VLT) have confirmed and revealed more detailed information on the azimuthal dips in TDs. For instance, another example of a warp related to a tilted inner disc is HD\,100453. \citet{benisty17} observed polarised scattered light using VLT/SPHERE at optical and Near-IR wavelengths, and found azimuthal dips interpreted as narrow shadows by \citet{min+17}. However, scattered light observations of shadowed outer rings often do not have radio counterparts. This is the case for HD\,100453 \citep{vanderplas+2019}, and also for warped disc HD\,143006 \citep{perezL+2018}, both of which show shadows in SPHERE polarised intensity with no radio counterparts in ALMA Band\,6. The absence of radio decrements under the shadows may be due to disc thermal radiation, which if optically thin \footnote{In the Planck or Rosseland sense.}, can smooth out the temperature decrement \citep{casassus_2019_cooling}. Thus, for a fixed dust population, such temperature decrements would be more conspicuous in massive discs.

\begin{figure*}
\begin{subfigure}{0.33\textwidth}
\includegraphics[width=\linewidth]{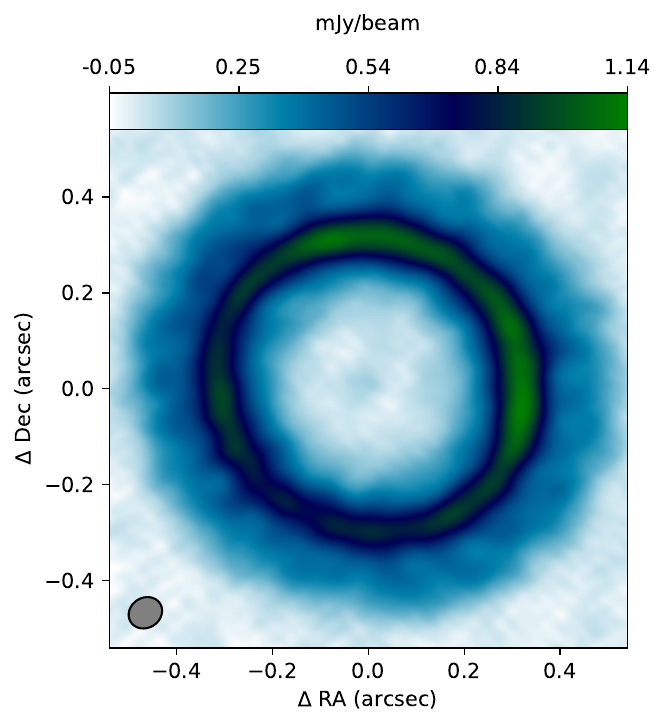}
\caption{ } 
\label{fig:b-fig}
\end{subfigure}\hspace*{0.5em}
\begin{subfigure}{0.33\textwidth}
\includegraphics[width=\linewidth]{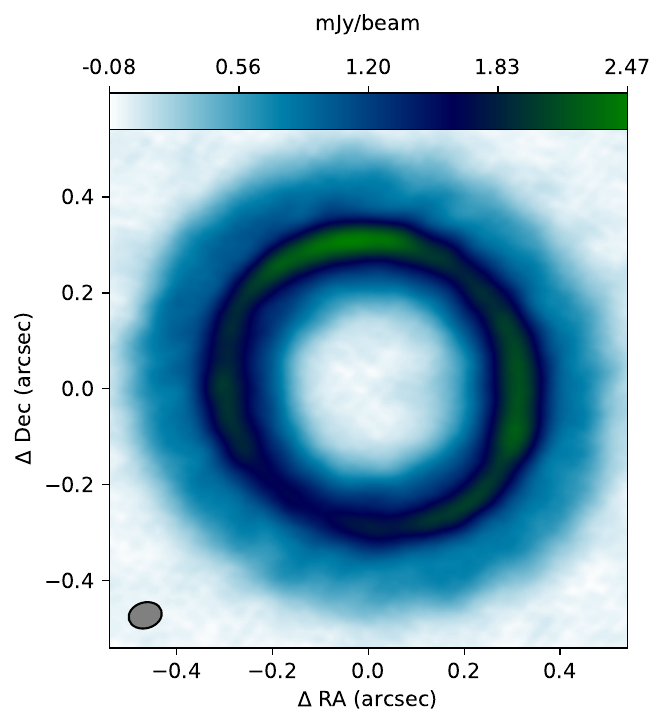}
\caption{ } 
\label{fig:c-fig}
\end{subfigure}\hspace*{0.5em}
\begin{subfigure}{0.33\textwidth}
\includegraphics[width=\linewidth]{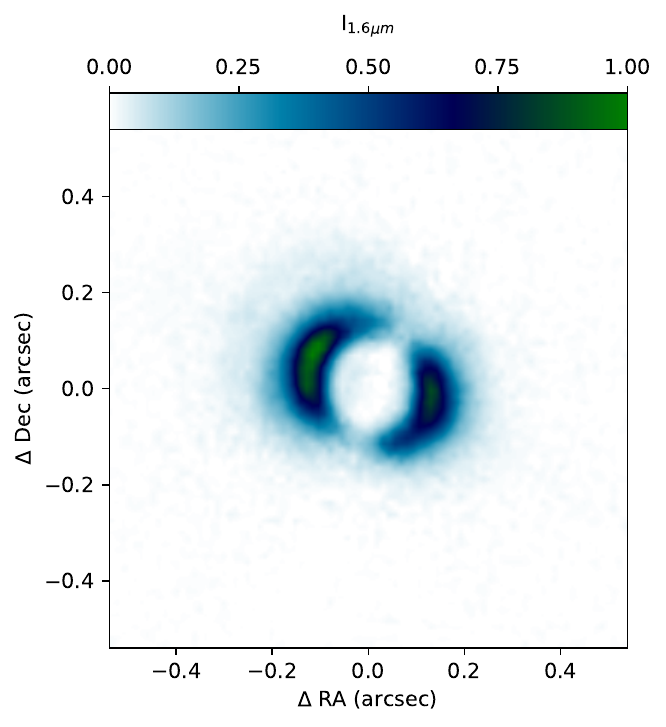}
\caption{ }
\label{fig:a-fig}
\end{subfigure}
\caption{\ref{fig:b-fig}: 230\,GHz continuum restored image. The grey ellipse shows the synthesized beam (72.4\,mas $\times$\,62.7\,mas for Briggs 1.0). \ref{fig:c-fig}: the 350\,GHz continuum restored image. The grey ellipse shows the synthesized beam (70.2\,mas\,$\times$\,53.8\,mas for Briggs 1.0). \ref{fig:a-fig}: $H$-band polarised intensity image, $Q_{\phi}(H)$.}
\label{fig:images}
\end{figure*}

An example of smoothed radio decrements due to radiative diffusion can be observed in DoAr\,44. DoAr\,44 (also known as WSB\,72, HBC\,268 or ROXs\,44) is a T-Tauri young stellar object (YSO) located in the L1689 region of the Ophiuchus dark cloud \citep{padgett08, andrews11} at a distance of 145.9\,$\pm$\,1.0\,pc \citep{gaia18}. It is catalogued as a pre-transition disc (PTD) as represented by its spectral energy distribution (SED) in \cite{cieza21}, with a SED slope $\alpha_{\mathrm{IR}}>$\,=\,$-$\,0.61. Here, PTDs are defined as Class\,II sources ($-$0.3\,$>\alpha_{\mathrm{IR}}>-$1.6) with an IR bright inner disc and a mid-IR SED decrement that indicates the existence of a gap. Also, DoAr\,44 is included in the multiplicity survey by \cite{zurlo20}, but no companion was detected.

Previous works on DoAr\,44, using SMA (Submillimeter Array) 340\,GHz continuum observations at an angular resolution of \SI{0.3}{\arcsecond}\citep{andrews09, andrews11} and ALMA 335\,GHz continuum with spectral observations \citep{vandermarel16} have detected a moderate $\sim$\,30\,au cavity in the disc. \citet[][hereafter, paper\,I]{casassus18} proposed a warped geometry inside the cavity by studying azimuthal decrements using ALMA 336\,GHz continuum observations, complemented with SPHERE/IRDIS (Infrared Dual-band Imager and Spectrograph) differential polarised imaging (DPI) and radiative transfer (RT) parametric models. A comparison with RT predictions suggests that the inner disc is tilted by $\xi$\,$\sim$\,30\,deg relative to the plane of the outer disc. However, the IRDIS data are affected by the coronograph, which borders the edge of the cavity. Paper\,I used aggressive deconvolution in order to improve the resolution of the ALMA data available at the time for their analysis, which suggested the presence of a radio dip associated to the northern shadow.

Here, we report new short and long-baseline (SB+LB) ALMA observations and a 3-year time difference follow-up SPHERE/IRDIS VLT observations of DoAr\,44. We aim to analyse the variability of the shadows generated by the warped inner disc over the outer disc, using multi-epoch ALMA Band\,6 and Band\,7 data and comparing follow-up SPHERE polarised intensity observations to the previous published epoch.

Section\,\ref{sec:obs} of this paper describes the ALMA and SPHERE observations, along with the data reduction. In Sec.\,\ref{sec:analysis} we describe the characteristics of the azimuthal intensity decrements, of the orientation of the inner disc and of the spectral index map between the radio bands, and Sec.\,\ref{sec:conc} concludes.

\begin{figure*}

\begin{subfigure}{0.47\textwidth}
\includegraphics[width=\linewidth]{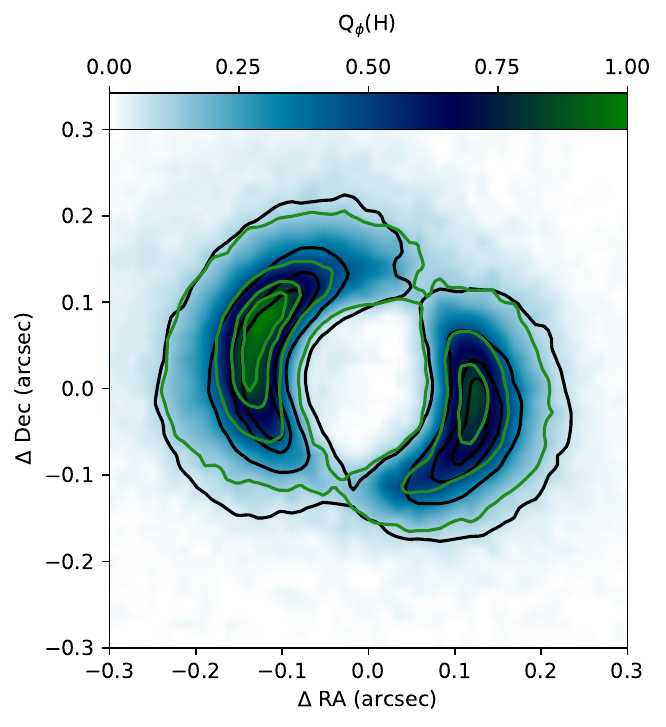}
\caption{ } 
\label{fig:2a-sphere}
\end{subfigure}\hspace*{0.5em}
\begin{subfigure}{0.47\textwidth}
\includegraphics[width=\linewidth]{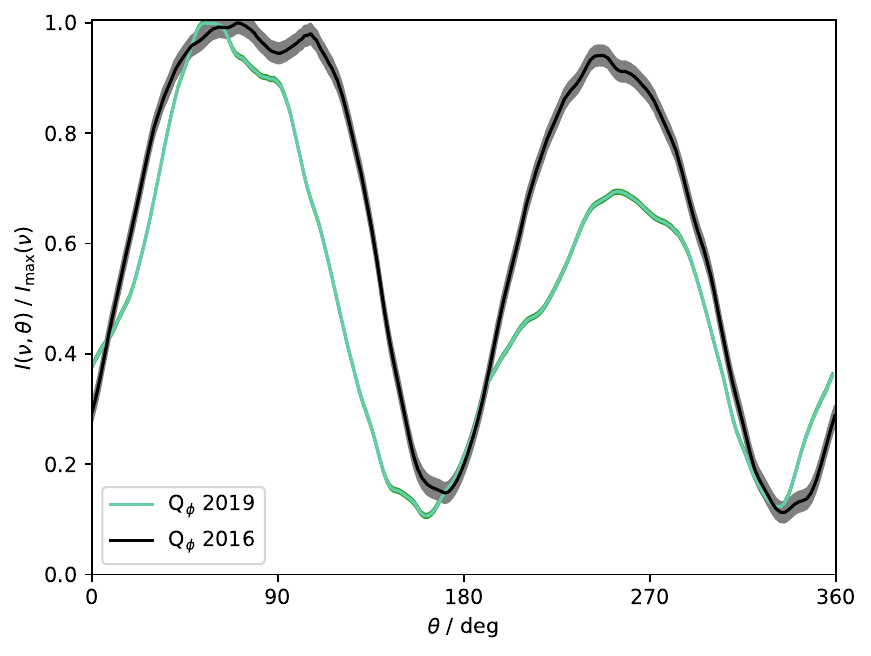}
\caption{ } 
\label{fig:2b-sphere}
\end{subfigure}
\caption{\ref{fig:2a-sphere}: $H$-band polarised intensity image, $Q_{\phi}(H)$, after unresolved polarisation subtraction. The green referential contours correspond to our $Q_{\phi}(H)$ intensity image, while the black contours correspond to the 2016 $Q_{\phi}(H)$ polarised intensity observations presented by paper\,I, both after unresolved polarisation subtraction. Both intensity contours are spaced at fractions of 0.13, 0.48, 0.74 and 0.87 times the peak intensity. \ref{fig:2b-sphere}: Ring intensity profiles extracted along constant radii (east of north): the green line corresponds to our $Q_{\phi}(H)$ observations after unresolved polarisation subtraction and the black line corresponds to the previous $Q_{\phi}(H)$ observations, from 2016, analysed in \citet{avenhaus_2018_sphere} and in paper\,I. The total height of the shaded areas correspond to $\pm$1$\sigma$.}

\label{fig:sphere_comparison}
\end{figure*}

\section{Observations and Data Reduction}
\label{sec:obs}
\subsection{ALMA observations} 
\label{sec:almadata}

The Ophiuchus DIsc Survey Employing ALMA \citep[ODISEA,][]{cieza19} LB survey was carried out during ALMA Cycle 6 in Band\,6 (230\,GHz), reaching resolutions of $\sim$\,3-5\,au (PID\,=\,2018.1.00028.S). The aim of this survey is to study the dust and gas components of a flux-limited population of discs, in a wide range of evolutionary stages, located in the Ophiuchus star-forming region at a distance of $\sim\,200$\,pc \citep{evans09, williams19}. These Band\,6 continuum observations of DoAr\,44 (bandwidth of 7.5\,GHz) were acquired in two epochs during July of 2019, and the data were calibrated using the ALMA Calibration Pipeline in CASA \citep[see][for details]{cieza21}. The continuum Band\,7 observations of DoAr\,44 (PID\,=\,2019.1.00532.S) were carried during May of 2021, in Cycle 7, with an on-source integration time of 43 minutes and a bandwidth of 4.688\,GHz. 
 
Self-calibration of the continuum data-sets was performed with the {\tt SNOW} package \citep[as in][]{casassus_22_selfcal}, which uses the {\tt uvmem} package \citep[][]{casassus_22_selfcal,  casassus21, casassus18} to replace the {\tt tclean} model in the CASA selfcal iterations \citep{casa}. The resulting 
peak signal to noise ratios (PSNR) in the restored images are $\sim$\,60 for Band\,6 and PSNR\,$\sim$\,100 for Band\,7, for Briggs weights with robustness parameter of 1.0. For comparison, the standard {\tt tclean} selfcal resulted in PSNR\,$\sim$\,49 in Band\,6 and PSNR\,$\sim$\,90 in Band\,7. Finally, the Band\,6 data-set was aligned to the Band\,7 data-set using the {\tt VisAlign} package \citep{casassus_22_selfcal}, but without scaling in flux.

Figure\,\ref{fig:b-fig} shows the reconstructed image for ALMA Band\,6, with an rms noise of $\sigma_{230\,\mathrm{GHz}}\,\sim\,$0.019\,mJy\,beam$^{-1}$ and a synthesized {\tt tclean} beam of 72.4\,mas\,$\times$\,62.7\,mas. Figure \ref{fig:c-fig} shows the reconstructed image for ALMA Band\,7, with an rms noise of $\sigma_{350\,\mathrm{GHz}}\,\sim$0.025\,mJy\,beam$^{-1}$ and a synthesized {\tt tclean} beam of 70.2\,mas\,$\times$\,53.8\,mas.

For the position of the star, we used \textit{Gaia} coordinates \citep{gaia18} of RA\,16:31:33.4638\,($\pm$\,0.0534\,mas), DEC\,$-$24:27:37.1582\,($\pm$\,0.0404\,mas), which were extrapolated to RA\,16:31:33.454\,, DEC\,$-$24:27:37.682\, by using a proper motion of ($-$6.101\,$\pm$\,0.128, $-$24.212\,$\pm$\,0.098)\,mas\,yr$^{-1}$. The center of the cavity (as calculated with {\tt MPolarMaps} in Sec.\,\ref{sec:profiles}), is shifted, with respect to the \textit{Gaia} coordinates by $\Delta$RA\,$=$\,1\,mas, $\Delta$DEC\,$=$\,6\,mas in Band\,6, and by $\Delta$RA\,$=$\,6\,mas, $\Delta$DEC\,$=$\,7\,mas in Band\,7, as presented in Table\,\ref{tab:pa_inc}. Given this, we can safely discard that this shift is due to a \textit{Gaia} pointing error ($\sim$0.15\,mas at 3$\sigma$). On the other hand, the accuracy of the ALMA data astrometry is regularly taken as $\sim$1/10 of the synthetic beam \citep{alma_book}, which translates into $\sim$7\,mas for both Band\,6 and Band\,7. This implies an ALMA pointing error of 21\,mas at 3$\sigma$, so the shifted cavity coordinates for both bands are within the errors.

\subsection{SPHERE observations}
\label{sec:spheredata}
We also used data from follow-up observations with SPHERE, the planet finder of the VLT. In this work, we use the 2019 data (preceded by 2016  096.C-0523(A) data, PI Avenhaus) from IRDIS \citep{dohlen08}, a differential imaging NIR camera (working in a range from 900\,nm to 2.3\,$\upmu$m), in the DPI mode, \citep[see, e.g.,][]{deboer2020}. The DoAr\,44 data were acquired with the IRDIS instrument on September 22 of 2019, and consisted of 60 frames in total. The observation was done in coronagraphic polarimetry mode. The 56 science frames (14 polarimetric cycles) were flanked with `flux' and `centering' frames. The flux frames are taken with the star displaced from the coronagraph and allow to estimate the flux of the star. Then, with the `N\_ALC\_YJH\_S' coronagraph (with a radius of 92.5 mas and an inner working angle, IWA, of $\sim$0.15\,arcsec) in place, the acquired star center frames allow a precise determination of the position of the star behind the coronagraph. There were no unexpected artifacts in the science frames and all the polarimetric cycles were successfully completed with a closed adaptive optics loop, thus, no data were discarded to produce the final images. As a result, these scattered light observations were conducted with a better seeing in comparison to the observations presented in \citet{avenhaus_2018_sphere} and in paper\,I. The seeing varied from 0.41$"$ at the beginning of the observations, had a peak at around 0.8$"$ and went down to $\sim$\,0.6$"$ at the end of the run. We also acquired data of DoAr\,44 using the Zurich Imaging Polarimeter (ZIMPOL), however, these observations were designed without considering a coronograph, resulting in lower quality images that were not considered for this work.

The SPHERE data reduction was performed with the publicly available pipeline {\sc irdap} \citep[IRDIS Data reduction for Accurate Polarimetry,][]{VanHolstein2020A&A...633A..64V} version 1.3.3. {\sc irdap} extracts the images of the left and right optical channels and performs the centering using the star center frames. For each measurement taken at half wave plate (HWP) angles (0, 45, 22.5, and 67.5) the left beam is subtracted from the right beam producing the Q+, Q-, U+ and U- (double difference) images. The single sum of these left and right images (double sum) is used to produce the total intensity images. The double difference images are used to obtain the Stokes Q, U and I images. Subsequently, {\sc irdap} corrects the Stokes images for instrumental polarisation using a detailed model of the instrument's optical path. Here we performed the analysis using the $H$-band polarised intensity image, $Q_{\phi}(H)$, as shown in Fig.\,\ref{fig:a-fig}.

\begin{figure*}
\begin{subfigure}{0.47\textwidth}
\includegraphics[width=\linewidth]{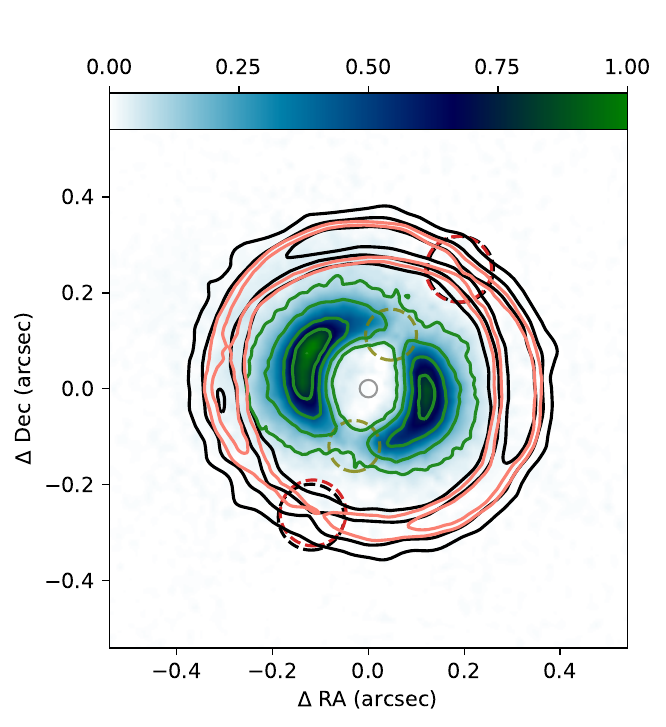}
\caption{ } 
\label{fig:3a-fig}
\end{subfigure}
\begin{subfigure}{0.47\textwidth}
\includegraphics[width=\linewidth]{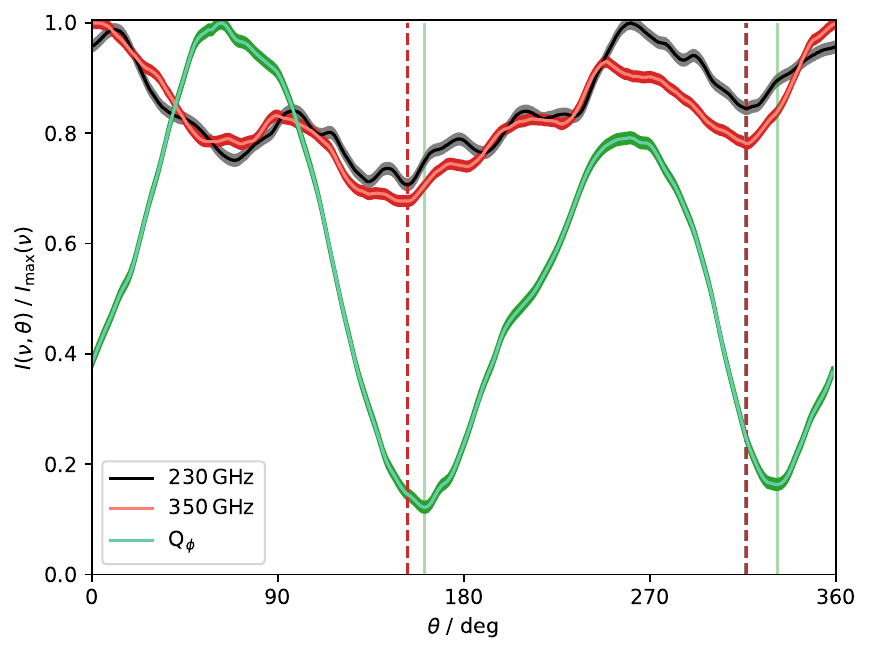}\caption{ } 
\label{fig:3b-fig}
\end{subfigure}
\caption{\ref{fig:3a-fig}: the 230\,GHz and the 350\,GHz restored {\tt uvmem} continuum in black and red contours, respectively, overlaid on the $Q_{\phi}(H)$-band image. The 230\,GHz intensity contours are spaced at fractions of 0.5, 0.6, 0.7, 0.8 and 0.9 times the peak intensity and the 350\,GHz contours are spaced at fractions of 0.68, 0.82 and 0.94 times the peak intensity. We also show in green the contours for $Q_{\phi}(H)$ at 0.1, 0.5, 0.7 and 0.9 times the peak. The grey circular marker at the center of the image indicates the position of the star (Sec.\,\ref{sec:profiles}). The dashed circular markers indicate the location of the decrements along projected circles: black for 230\,GHz and 350\,GHz, and green for $Q_{\phi}(H)$. \ref{fig:3b-fig}: Ring intensity profiles extracted along constant radii. The green line corresponds to $I_{H}$($\theta$), the red line corresponds to $I_{\textrm{350 GHz}}$($\theta$) and the grey line corresponds to $I_{\textrm{230 GHz}}$($\theta$). The dashed grey and red lines mark the location of the decrements for 230\,GHz and 350\,GHz, respectively, and the solid green lines mark the location of the decrements for $Q_{\phi}(H)$. Note that the angle $\theta$ in \ref{fig:3b-fig} increases in the east of north direction. Also note that the $I_{H}$($\theta$) profile was integrated along a range of radii (0.08\,arcsec $< r <$ 0.22\,arcsec) in order to obtain a better signal. The total height of the shaded areas correspond to 1$\sigma$.}
\label{fig:decrements}
\end{figure*}

A crucial step in the reduction is the subtraction of the halo of polarised signal associated with the unresolved signal within the central PSF, the so-called "unresolved" polarisation. To remove this component we selected an annular region devoid of disc signal centered on the star with a radius of 40 to 90 pixels. The degree of linear polarisation measured from the unresolved central source in DoAr\,44 is 1.6$\pm$0.4\%, the polarisation angle (AoLP) of the central source is AoLP$_{\mathrm{central}}$\,=\,19.97$\pm$7.25\,deg, and the uncertainty is the standard deviation of the unresolved polarisation signal during the polarimetric cycles. This uncertainty is caused by variations in the  atmospheric conditions and is higher than the statistical uncertainty and the accuracy of the Mueller matrix model used to correct the instrumental polarisation. 

An interesting feature to point out is that the locations of the intensity decrements change after the unresolved polarisation subtraction. This effect was seen for different locations of the signal-devoided annular region. The polarised light halo that is associated to the unresolved signal most likely has an asymmetric pattern due to unresolved tilted inner disc around the star. This pattern contributes to a variability in the contrast of the intensity distribution \citep[as also observed by][using SPHERE/IRDIS data of GG\,Tauri\,A]{keppler20}, so, when subtracted, the location of the decrements are slightly affected. This is expected as a result of correcting for the shadowing produced by the inner interstellar (IS) polarisation. 

In Figure\,\ref{fig:2a-sphere} we compare our SPHERE/IRDIS $Q_{\phi}(H)$ observations with the SPHERE/IRDIS 2016 $Q_{\phi}(H)$ observations described by \citet{avenhaus_2018_sphere} (project 0.96.C-0523(A)) and presented in paper\,I. Note that we subtracted the unresolved polarisation from both data-sets using {\sc irdap} and considering the parameters described above. As confirmed by the profiles in Figure\,\ref{fig:2b-sphere}, the location of the north-western dip varies slightly between both data-sets, however, they show different locations for the south-eastern decrement. There are also variations in the peak intensities between epochs. This discrepancies will be further addressed in Section\,\ref{sec:warpgeo}.

\section{Analysis}
\label{sec:analysis}

\subsection{Azimuthal profiles}
\label{sec:profiles}
We trace the location of the decrements in the ring around DoAr\,44 with the {\tt MPolarMaps} package, described in \citet{casassus21}. This allows us to extract the ring intensity profile $I^{\circ}$($\theta$) as a function of azimuth $\theta$ along a circle that best approximates the disc ring, by minimizing the dispersion in the radial profile of the continuum and under the assumption of axial symmetry. As a result, we get the best fits values for the position angle (PA), the inclination ($i$) and the stellar offsets.

As shown in Table\,\ref{tab:pa_inc}, the 230\,GHz ALMA data gives an inclination $i$\,=\,21.7\,$\pm$\,0.3\,deg and a PA\,=\,59.9\,$\pm$\,0.8\,deg, and the offsets from the nominal stellar position are $\Delta$RA\,=\,1\,mas and $\Delta$DEC\,=\,6\,mas. Similarly, the optimization for the 350\,GHz ALMA data gives an $i$\,=\,21.9\,$\pm$\,0.2\,deg and a PA\,=\,61,6\,$\pm$\,0.5\,deg, with offsets from the nominal stellar position as $\Delta$RA\,=\,6\,mas and $\Delta$DEC\,=\,7\,mas. Since the $Q_{\phi}(H)$ continuum does not show a continuous disc geometry due to its pronounced decrements, we fitted a continuous ellipse to the emission and used the result to run the {\tt MPolarMaps} optimization. This gives an $i$\,=\,24.4\,deg and a PA\,=\,49.7\,deg for the $Q_{\phi}(H)$ data, with offsets from the nominal stellar position as $\Delta$RA\,=\,2\,mas and $\Delta$DEC\,=\,2\,mas. Note that the measurement of $I^{\circ}$($\theta$) depends on the choice of origin, but, as explained in Section\,\ref{sec:almadata}, the offset between the stellar position and the center of the cavity is within the ALMA and \textit{Gaia} pointing errors. Given this, we can safely center the cavity to \textit{Gaia} coordinates, as presented in Figure\,\ref{fig:3a-fig}. 

\begin{table}
	\centering
	\caption{Position angle and ring inclination values for DoAr44 from the polar optimization.}
	\begin{threeparttable}
	\begin{tabular}{lccc} 
		\hline
		  & 230\,GHz & 350\,GHz & $Q_{\phi}(H)$ fit\tnote{a}\\
		\hline
    	Position Angle (PA)\tnote{b} & 59.2\,$\pm$\,0.8 & 61.6\,$\pm$\,0.5 & 49.7$\pm$\,1.9 \\
    	
		Inclination ($i$) & 21.7\,$\pm$\,0.3 & 21.9\,$\pm$\,0.2 & 24.4$\pm$\,0.8 \\
		RA offset (mas) & 1\,$\pm$\,0.1 & 6\,$\pm$\,0.1 & 2$\pm$\,0.1 \\
		DEC offset (mas) & 6\,$\pm$\,0.1 & 7\,$\pm$\,0.1 & 2$\pm$\,0.1 \\
		\hline
	\end{tabular}
	\begin{tablenotes}
	\item[a] $Q_{\phi}(H)$ observations do not show a continuous disc, which is required for a correct polar optimization, thus, we performed a Gaussian fit along the emission peak in order to calculate the disc PA and inclination.
    \item[b] East of North.
    \end{tablenotes}
    \end{threeparttable}
    \label{tab:pa_inc}
\end{table}

\begin{table*}
\caption{Summary of results from the optimization for the intensity decrements in the 230\,GHz and 350\,GHz continuum and $Q_{\phi}(H)$.} 

\centering 
\begin{threeparttable}
\begin{tabular}{c  ccc  ccc}
\toprule
  & \multicolumn{3}{c}{Decrement 1 (south)} &
\multicolumn{3}{c}{Decrement 2 (north)} \\ \cline{2-7}
& 230\,GHz & 350\,GHz & $Q_{\phi}(H)$ & 230\,GHz & 350\,GHz & $Q_{\phi}(H)$   \\
\midrule

I$_{\mathrm{D}}$/I$_{\mathrm{Peak}}^\mathrm{a}$ & 98$\pm$1$\%$ & 81$\pm$1$\%$ & 11$\pm$0.1$\%$ & 83$\pm$1$\%$ & 89$\pm$1$\%$ & 13$\pm$0.1$\%$   \\

$\theta_{\mathrm{sky}}$[$\circ$]$^\mathrm{b}$ & 152.9 & 152.9 & 160.9 & 316.3 & 316.9 & 331.8  \\ 

r$_{\mathrm{star}}$[arcsec]$^\mathrm{c}$ & 0.29 & 0.28 & 0.12 & 0.32 & 0.31 & 0.12 \\

\bottomrule
\end{tabular}
\begin{tablenotes}
\item[a,b]$^\mathrm{a}$ Minimum intensity over peak along the ring.
\item[a,b]$^\mathrm{b}$ Position Angle (location) of the decrement on the plane of the sky East of North. The errors for these values are calculated as 1/10 of the resolutions, which are in the order of 10$^{-6}$.
\item[a,b]$^\mathrm{c}$ Stellocentric separation of the decrements on the sky.
\end{tablenotes}
\end{threeparttable}
\label{tab:summary}
\end{table*}

The {\tt MPolarMaps} also records the intensity profiles as a function of azimuth for the radial maxima along the ring, which lets us estimate the location of the intensity decrements. In Figure\,\ref{fig:3b-fig} we show the ring intensity profiles for all data-sets and the positions of the radio/IR intensity decrements along the projected circles. These positions, indicated by thin lines, correspond to the minima in $I^{\circ}$($\theta$). Table\,\ref{tab:summary} shows the results for the optimization: the PA of the decrement (East of North) as viewed on the sky ($\theta_{\mathrm{sky}}$) and the stellocentric distance to the decrements on the sky (r$_{\mathrm{star}}$).

Table\,\ref{tab:summary} shows that the PAs for both decrements at 230\,GHz, 350\,GHz and $Q_{\phi}(H)$ are closely coincident. Also, it is interesting to note that the mm-continuum, towards the north-east in Figure\,\ref{fig:3a-fig}, is nearly coincident with the maximum in the $Q_{\phi}(H)$ emission, as also seen in Figure\,\ref{fig:3b-fig}. We find similar results in paper\,I, where they tested parametric models in order to analyse the characteristics of the disc. Their results suggested that a tilted inner disc provided the simplest explanation for the presence of the decrements in DoAr\,44. An inner disc has been detected with ALMA in $^{12}$CO emission by \citet{antilen+23}.

\begin{table}
	\centering
	\caption{Observed and calculated values for the orientation of the inner disc in DoAr44 using the idealized geometrical argument.}
	\begin{threeparttable}
	\begin{tabular}{lc} 
		\hline
		Observed parameters in $Q_{\phi}(H)$ &  \\
		\hline
    	PA & 161.7\,deg \\
		Angular separation ($\omega$) & 170.9\,deg \\
		Stellar offset ($b$) & 0.14 \\
        \hline
    	Calculated parameters for the inner disc  &  \\
		\hline
		\vspace{0.1cm}
		Inclination ($i_1$) & 26.4\,$^{+5.6}_{-3.5}$\,deg \\
		\vspace{0.1cm}
		PA ($\phi_1$) & 110.7\,$^{+12.7}_{-13.9}$\,deg \\
		\vspace{0.1cm}
		Scale Height ($h$) & 0.014\,$^{+0.006}_{-0.004}$ \\
		Relative Inclination ($\xi$) & 21.4\,$^{+6.7}_{-8.3}$\,deg \\
		
		\hline
	\end{tabular}
    \end{threeparttable}
    \label{tab:warpgeo}
\end{table}

\subsection[]{Radio/IR intensity decrements} 
\label{sec:decrements}
 
The $Q_{\phi}(H)$ image (Fig.\,\ref{fig:a-fig}) shows a ring divided into bipolar arcs separated by deep intensity decrements, while the 230\,GHz and 350\,GHz continuum images (Figs.\,\ref{fig:b-fig} and \ref{fig:c-fig}) show a smoother and approximately face-on projected circle. However, it is interesting to note the contrast between the mm-continuum and the IR in Fig.\,\ref{fig:3b-fig}, which shows that the decrements are much more prominent in the IR data. Also, the radio decrements of the disc appear to be shifted ahead of the IR decrements in the direction West of North (clockwise). The angular separations between the radio and IR dips on the plane of the disc ($\eta_{\mathrm{shift}}$) are $\eta_{\mathrm{shift}}$\,=\,14.95\,deg for the northern decrements, and $\eta_{\mathrm{shift}}$\,=\,7.92\,deg for the southern decrements. As the dust and gas enter the shadows, we would expect to see the dust temperature decrements shifted with respect to the IR scattered light shadows \citep{casassus_2019_cooling}. Provided a clockwise direction of rotation, the observed offset between the radio and IR shadows could be attributed to a thermal lag, and it is interesting to point out that this clockwise offset, where the radio decrements lead, is observed in different epochs (in paper I and in this work). However, a more detailed analysis with simultaneous multi-frequency imaging could provide a better grip on such thermal lag. We also note that the radio decrements in the south of the disc cannot be defined with high precision compared to the northern decrements: the disc appears to be less massive and narrower in the location of the southern decrements (as also proposed by \cite{casassus_2019_cooling}), hence with a lower optical depth to radiation diffusion, which smooths the temperature decrements. This will be further analysed in Sections\,\ref{sec:warpgeo} and \ref{sec:discmass}.


\subsection{Warped inner disc}
\label{sec:warpgeo}

Shadows in a transition disc can be interpreted as caused by a misaligned inner disc with respect to the outer disc \citep[e.g.][]{marino15, benisty17}. The scattered light $Q_{\phi}(H)$ image of DoAr\,44 clearly reveals the location of two shadows on the outer ring (Figs.\,\ref{fig:a-fig} and \ref{fig:decrements}) Also, Figure\,\ref{fig:sphere_comparison} shows a difference between the \citet{avenhaus_2018_sphere} $Q_{\phi}(H)$ observations and our $Q_{\phi}(H)$ observations. Note that the unresolved polarised emission was subtracted from both data-sets using the same parameters in {\sc irdap} (Sec.\,\ref{sec:spheredata}), but the observing conditions in each epoch could have an effect on the applied correction. However, the unmatched locations of the dips between the two infrared profiles would point toward a morphological change in the location of the dips between epochs, which appears to be more significant for the southern dip. Changes in the location and morphology of intensity dips have been previously identified in other discs, for instance, in J1604, using multi-epoch scattered light observations \citep{pinilla+18}. In the case of DoAr\,44, the variability in the morphology and the location of the dips is most likely related to an illumination effect due to a shift in the orientation of the tilted inner disc. We analyse this interpretation by using two approaches: an idealized geometrical model for the location of the shadows and a qualitative analysis with radiative transfer (RT) predictions. 

\begin{figure*}
\begin{subfigure}{0.47\textwidth}
\includegraphics[width=\linewidth]{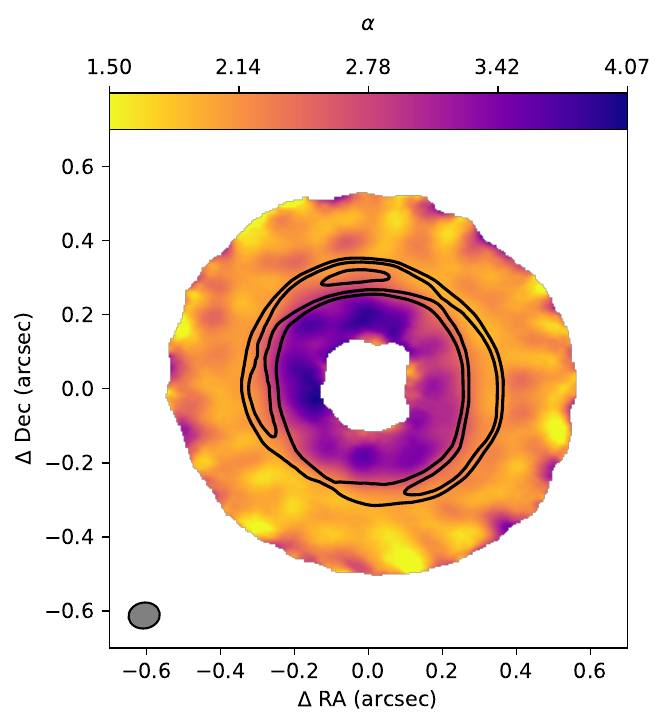}
\caption{ } 
\label{fig:4a-fig}
\end{subfigure}
\begin{subfigure}{0.47\textwidth}
\includegraphics[width=\linewidth]{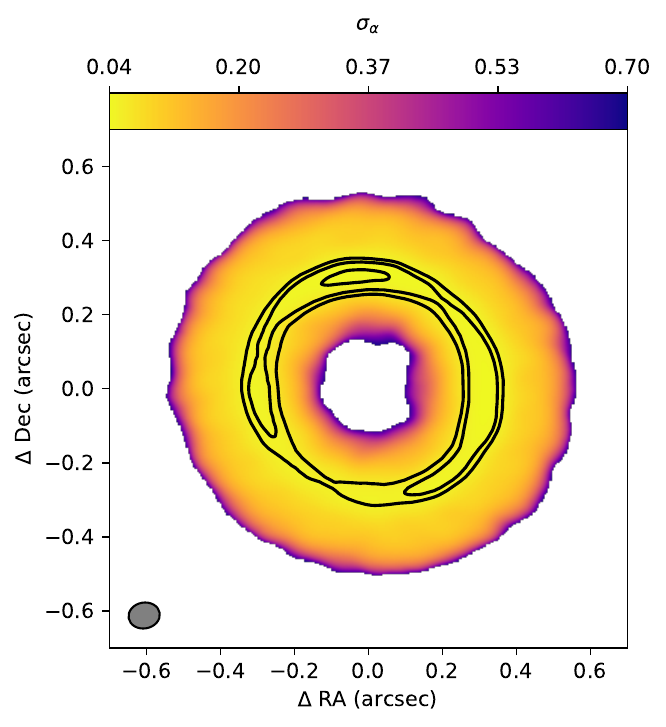}\caption{ }
\label{fig:4b-fig}
\end{subfigure}
\caption{\ref{fig:4a-fig}: Spectral index map of DoAr\,44, computed between 230\,GHz and 350\,GHz. We also show the 350\,GHz intensity profiles in black contours, at 0.67, 0.77 and 0.95 times the peak, in order to highlight the location of the decrement. The grey ellipse shows the synthesized beam (84.9\,mas\,$\times$\,69.9\,mas). \ref{fig:4b-fig}: Spectral index error map with 350\,GHz intensity profiles in black contours, at 0.67, 0.77 and 0.95 times the peak.} 
\label{fig:specind}
\end{figure*}

\subsubsection{Idealized geometrical model}
\cite{min+17} propose an algorithm that considers the location of the observed shadows as the intersection of the inner disc midplane with a perfectly circular outer disc. This scheme relates the inner disc orientation (this is, inclination ($i_{1}$) and PA ($\phi_{1}$) on the sky), to the scale height ($h$) of the outer disc, the PA of the observed shadows (PA) and their subtended angle relative to the star ($\omega$), and the stellar offset ($b$, the intercept of the inner disc PA with the North). In practice, however, some uncertainties arise: the fact that the observed $Q_{\phi}(H)$ cavity is not perfectly circular makes it difficult to determine its center with precision. Also, the exact locations of the center of the shadows, this is, the decrement minimum, might not necessarily be in the midplane of the inner disc. Taking these caveats into consideration, we calculated the inner disc orientation by considering the observed values presented in Table\,\ref{tab:warpgeo}. These give us an inner disc orientation of $i_{1}$\,=\,26.4\,$^{+5.6}_{-3.5}$\,deg, $\phi_{1}$ \,=110.7\,$^{+12.7}_{-13.9}$\,deg and $h$\,=0.014\,$^{+0.006}_{-0.
004}$, with uncertainties corresponding to the 1$\sigma$ confidence intervals. We find that the PA value for $Q_{\phi}(H)$ (Table\,\ref{tab:summary}) is significantly different than that of the inner disc. These results provide support to the interpretation of the shadows being caused by a misalignment between the inner and outer discs. \cite{bouvier+20} derived similar results for the orientation of DoAr\,44's inner disc: they fitted the continuum visibilities of 2019 VLT/GRAVITY observations and calculated an inclination of $i_{1}$\,=\,34\,$\pm$\,2\,deg and a PA of $\phi_{1}$ \,=140\,$\pm$\,3\,deg. Since polarisation vectors are azimuthally oriented in discs, the AoLP can be either parallel or perpendicular to the direction of the disc PA \citep[e.g.,][]{stolker+17}. It is interesting to note that the polarisation angle of the central source, AoLP$_{\mathrm{central}}$\,=19.97$\pm$\,7.25\,deg, is perpendicular to the inner disc PA, $\phi_{1}$ \,=110.7\,$^{+12.7}_{-13.9}$\,deg. Also, in Fig.\,\ref{fig:decrements} we see that the radio intensity decrements lead in the direction west of north, both in the northern and southern regions. Considering the geometrical approach shown in \citet[][Fig.\,1]{min+17}, if the azimuthal shift was only related to projection effects due to the emission heights, we would see the northern and southern decrements shifted opposite relative to rotation.

The orientation values we calculated from the idealized geometrical method for DoAr\,44's warped inner disc correspond to a relative orientation of $\xi$\,=\,21.4\,$^{+6.7}_{-8.3}$ deg with respect to the outer disc. We can compare this orientation of the inner disc with the results of \citet{bohn+22}, who fit parametric models to the 2019 VLT/GRAVITY visibilities in order to derive the geometry of the inner disc, while the geometry of the outer disc was derived from fitting ALMA velocity maps. They  estimated two misalignment angles between the inner and outer discs, considering that, in the GRAVITY data, it is not possible to distinguish  which side of the
inner disc is closer to the observer: $\xi_{1}$\,=\,27\,$\pm$\,9\,deg, which is consistent with our result, and $\xi_{2}$\,=\,39\,$\pm$\,9\,deg. It is interesting to note that their PA and $i$ values, both for the outer and inner disc, are also in agreement with our PA and $i$ results.

\subsubsection{Radiative transfer predictions}
Our relative orientation of $\xi$\,=\,21.4\,$^{+6.7}_{-8.3}$ deg can be compared with the RT predictions for inner disc tilts presented in paper\,I. If we examine Figure\,4 in paper\,I, we find that our 2019 $Q_{\phi}(H)$ observations are better reproduced by a misalignment of $\xi$\,$\sim$\,20\,deg, which is in excellent agreement with $\xi$\,=\,21.4\,$^{+6.7}_{-8.3}$. On the other hand, a relative inclination of $\xi$\,$\sim$\,30\,deg better accounts for $Q_{\phi}(H)$ images from the 2016 data. Also, the azimuthal profiles for both epochs (Figure\,\ref{fig:2b-sphere}) are consistent with the RT predicted profiles in paper\,I (Fig.\,7), for the aforementioned $\xi$ values. It is interesting to note that the difference in the relative inclinations between epochs also translates into variations in the location of the shadows and in the relative brightness of the arcs between the shadows. The RT predictions for $\xi$\,=\,20\,deg result in broader decrements compared to the predictions for $\xi$\,=\,30\,deg. These effects are also observed comparing the azimuthal profiles for both $Q_{\phi}(H)$ epochs in Fig.\,\ref{fig:2b-sphere}, which implies a morphological variation, particularly, in the locations of the southern decrement and the western arc.

\subsection{Spectral index map}
\label{sec:specind}

If due to shadowing, we would expect a drop in the dust temperature and optical depth in the location of the intensity decrements. In order to analyse this, we compute the spectral index map between 230\,GHz and 350\,GHz, as 

\begin{ceqn}
\begin{align}
\alpha\,=\,\frac{\log(I_{230\,\mathrm{GHz}}/I_{350\,\mathrm{GHz}})}{\log(230\,\mathrm{GHz}/350\,\mathrm{GHz})},
\label{eq:specind}
\end{align}
\end{ceqn}

and the error map as

\begin{ceqn}
\begin{align}
\sigma_{\alpha}\,=\,\frac{\sqrt{(\sigma_{230\,\mathrm{GHz}}/I_{230\,\mathrm{GHz}})^2 + (\sigma_{350\,\mathrm{GHz}}/I_{350\,\mathrm{GHz}})^2}}{\log(230\,\mathrm{GHz}/350\,\mathrm{GHz})}.
\label{eq:specinderror}
\end{align}
\end{ceqn}
We smoothed the 230\,GHz image to obtain the same synthesized (elongated) beam as that of a Briggs 2.0 weighting scheme for 350\,GHz (84.9\,mas\,$\times$\,69.9\,mas), using the previously aligned images (Sec.\,\ref{sec:almadata}). It is interesting to note that the smoothed 230\,GHz emission does not show dips as clear as the  350\,GHz emission (Fig.\,\ref{fig:c-fig}). This could be due to the fact that the 350\,GHz emission is more optically thick than the 230\,GHz emission, thus, being sensitive to temperature only.

The $\alpha$ map was computed for intensities with errors $\sigma_{\alpha}$\,<\,0.7, which translates into intensities greater than 0.27\,mJy in Band\,6 (14\,$\sigma$) and 0.67\,mJy in Band\,7 (26\,$\sigma$). Note that a calibration uncertainty of 2.5\,$\%$ for Band\,6 and 5\,$\%$ for Band\,7 \citep{alma_book} translates into a $\pm$\,0.13 uncertainty in the absolute value of $\alpha$ (using error propagation, such that $\sigma_{\alpha}$\,=\,0.056\,/\,ln(350\,GHz\,/\,230\,GHz), but the values in the error map in Fig.\,\ref{fig:4b-fig} do not consider this systematic errors since these are uniform across the map and we are mostly interested in relative changes.

\begin{figure}
\includegraphics[width=\linewidth]{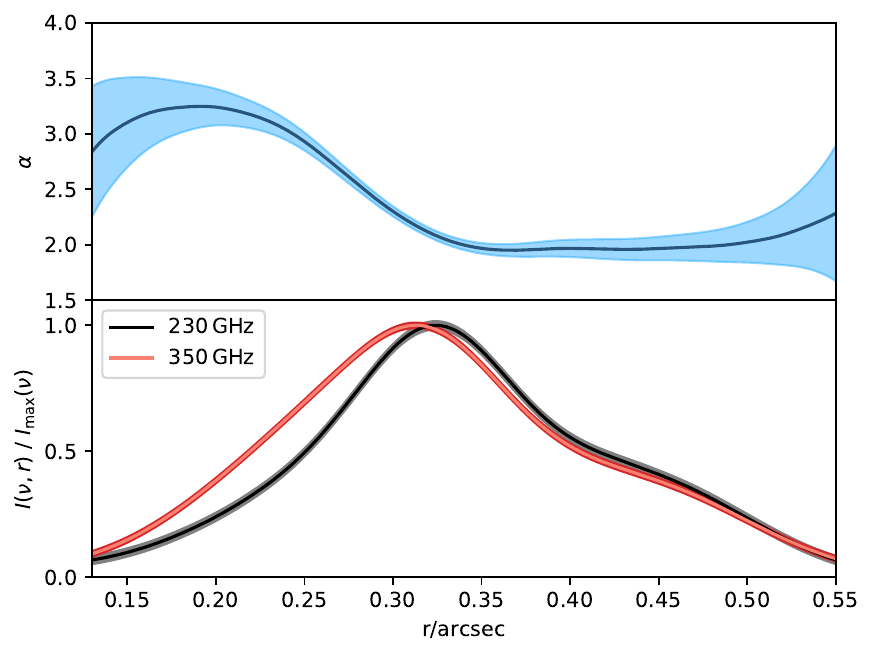}
\caption{Radial profile of the spectral index derived from the radial profiles of the 230\,GHz and 350\,GHz continuum following Eq.\,\ref{eq:specind}. The shaded regions trace the radial profile errors. The profile is derived following the radial range that corresponds to the values presented in Fig.\,\ref{fig:4a-fig} (0.13\,arcsec\,<\,$r$\,<\,0.55\,arcsec).}
\label{fig:radial_specind}
\end{figure}

In Fig.\,\ref{fig:specind}, we find spectral index values $\alpha$\,$\sim$\,2. The inner cavity (0.13\,arcsec\,$<$\,$r$\,$<$\,0.25\,arcsec, within the black intensity profiles in Fig.\,\ref{fig:specind}) shows $\alpha$ values much higher than $\sim$\,2, as also seen in Fig.\,\ref{fig:radial_specind}. This suggests an optically thin emission due to the presence of very small grains, most likely as a consequence of grain destruction by stellar radiation. The $\alpha$\,$\sim$\,2 values along the rest of the disc could either correspond to an optically thick emission, or suggest a shallow dust opacity index due to larger grains \citep[$\beta \sim$\,0, e.g.][]{testi14}. Furthermore, this is similar to the structure of HD\,142527 in \cite{casassus15b} who reported a drop in the intensity of the outer disc using ATCA and ALMA observations from 34\,GHz to 700\,GHz. We do not find distinct azimuthal spectral variations in the location of the decrements, where $\alpha$\,$\sim$\,2 suggests optically thick emission (Fig.\,\ref{fig:specind}), however, the dips in the infrared profile (Fig.\,\ref{fig:decrements}) point towards temperature decrements due to shadowing. We also calculated $\alpha$\,$>$\,$-$2 to be the lower limit for the star's spectral index at 1$\sigma$ intensities for Bands\,6 and 7.

It is particularly interesting that the spectral index values at 0.13\,arcsec\,<\,$r$\,<\,0.25\,arcsec are $\alpha$\,$\sim$\,3.3, as seen in Fig.\,\ref{fig:radial_specind}. This radial range corresponds to the inner cavity in Bands\,6 and 7, where a faint dust pedestal can be observed at the edge of the inner cavity (Figs.\,\ref{fig:b-fig} and \ref{fig:c-fig}). The spectral index values in the inner cavity are typical for the interstellar medium (ISM), therefore, considering a dust opacity index $\beta$\,=\,1.7 \citep[the ISM value,][]{draine2006}, this can be interpreted as optically thin emission and a population of small dust grains \citep[$a_{\mathrm{max}}$\,$<<$\,100\,$\mu$m,][]{sierra20} filling the cavity. This is consistent with the absence of intensity decrements in the inner cavity, which are smoothed out by optically thin thermal radiation from the dust pedestal. Also, it is intriguing that this high spectral index values are not seen towards $r$\,=\,0.5\,arcsec (Fig.\,\ref{fig:radial_specind}), where we observe an extended dust halo outside of the radial dust trap in both Bands\,6 and 7 (Figs.\,\ref{fig:b-fig} and \ref{fig:c-fig}). The $\alpha$\,$\sim$2-2.5 along the dust halo implies the presence of larger dust grains compared to the inner cavity. However, further multi-wavelength data and modelling is required to resolve these ambiguities in DoAr\,44.

\subsection{Disc gas mass estimates}
\label{sec:discmass}
The gas mass in discs is an important condition for the formation of planets, and in the case of shadowed transition discs, the cooling of the gas as it crosses the shadows can result on constraints on its mass. In this sense, \citet{casassus_2019_cooling} presented a 1D model that relates the outer disc surface density ($\Sigma_g$) to the depth of the temperature profiles ($T(\phi)$) due to radiation stemming from the central star and the inner disc. The model considers a shadow moving relative to the gas at a retrograde Keplerian speed, a fixed disc structure described by its radius ($R$) and thickness ($H$), and a standard dust population, with a maximum grain size ($a_{\mathrm{max}}$), the gas-to-dust mass ratio ($f_{\mathrm{gd}}$) and the grain filling factor ($f$). Hence, the free parameters governing the gas temperature are the observed angular shifts between the radio and IR decrements, $\eta_{\mathrm{shift}}$, and the gas surface density, $\Sigma_g$. Note that here we define $\eta_{\mathrm{shift}}$ on the plane of the disc, as opposed to paper\,I that defined $\eta_{\mathrm{shift}}$ on the plane of the sky.

In the case of DoAr\,44, the observed angular shifts between the radio and IR decrements, described in Section\,\ref{sec:decrements}, can be explained by a thermal lag between the shadowed dust and gas, provided that the outer disc has a retrograde direction of rotation. We computed the $T(\phi)$ profiles considering a maximum grain size of $a_{\mathrm{max}}$\,=\,1\,cm, which is consistent with the optically thick $\alpha \sim$\,2 values we found for the outer disc (Figs.\,\ref{fig:specind} and \ref{fig:radial_specind}). We also use the established gas-to-dust mass gas ratio for the diffuse ISM, $f_{\mathrm{gd}}$\,=\,100,  and a filling factor for compact grains $f$\,=\,1. These parameters are presented in Table\,\ref{tab:mass_parameters}.

\begin{figure}
\includegraphics[width=\linewidth]{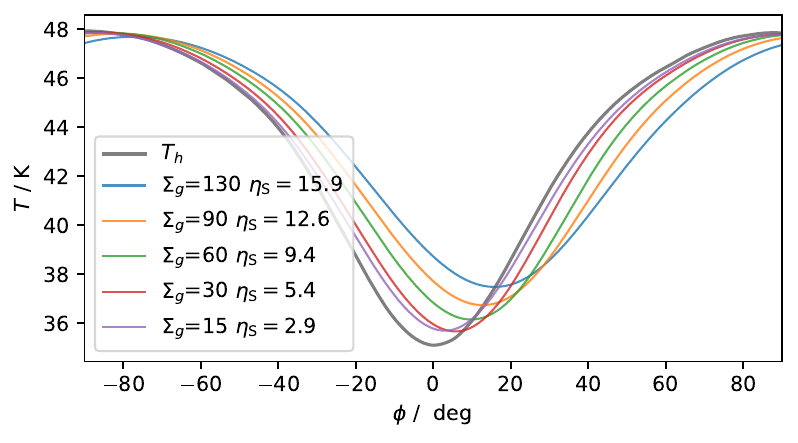}
\caption{Profiles $T(\phi)$ that approximate DoAr\,44. The curves were computed for $a_{\mathrm{max}}$\,=\,1\,cm, $f_{\mathrm{gd}}$\,=\,100 and $f$\,=\,1. The direction of the gas Keplerian rotation is towards $+ \phi$ }
\label{fig:disc_mass}
\end{figure}

Figure\,\ref{fig:disc_mass} shows the resulting $T(\phi)$ profiles for different values of $\eta_{\mathrm{shift}}$ and $\Sigma_g$. The angular shift between the northern decrements is $\eta_{\mathrm{shift}}$\,=\,14.95\,deg, which corresponds to an interpolated surface density of $\Sigma_g$\,=\,117\,$\pm$\,10 g/cm$^2$, and the southern decrements have an angular shift of $\eta_{\mathrm{shift}}$\,=\,7.92\,deg, which corresponds to an interpolated surface density of $\Sigma_g$\,=\,48\,$\pm$\,10 g/cm$^2$. A lower value of $\Sigma_g$ means that the location of the southern dip is optically thinner to thermal radiation, which is consistent with the southern decrement being smoother and broader (as seen in Figure\,\ref{fig:3b-fig}), as a result of thermal diffusion for the considered disc rotation. The contrast ratio between the estimated intensity decrements along the outer ring gas surface density is $f_r$\,=\,2.4\,$\pm$\,0.5, a value that is consistent with the presence of a large crescent of continuum sub-mm emission in the disc \citep{zhu_bauteau_16}, as in, e.g., LkH$\alpha$330 \citep{isella13}, HD\,135344B \citep{vandermarel15} and MWC\,758 \citep{casassus19_mwc758}. Also, it is interesting to note that, in the context of a lopsided vortex, the southern decrement may correspond to  significantly lower optical depths in the ALMA continuum. Yet the spectral index in the location of both decrements remains $\alpha \sim$2 (Fig.\,\ref{fig:specind}). If the southern decrement is optically thin, perhaps dust trapping in DoAr\,44 is efficient only for the denser large grains, while a population of fluffy large grains would account for spectral index of the continuum halo and the southern decrement. 

\begin{table}
	\centering
	\caption{Parameters for the disc mass calculation.}
	\begin{threeparttable}
	\begin{tabular}{lc} 
		\hline
		  Parameter & Value \\
		\hline
    	Angular shift between radio and IR dips ($\eta_{\mathrm{shift}}$) & north: 14.95\,deg \\
    	   & south: 7.92\,deg   \\
		Maximum grain size ($a_{\mathrm{max}}$) & 1 cm \\
		Gas-to-dust mass ratio ($f_{\mathrm{gd}}$) & 100 \\
		Filling factor ($f$) & 1 \\
		\hline
	\end{tabular}

    \end{threeparttable}
    \label{tab:mass_parameters}
\end{table}

\section{Conclusions}
\label{sec:conc}

We report ALMA 350\,GHz observations of DoAr\,44 with a linear resolution of $\sim$10\,au, as well as new differential polarised SPHERE/IRDIS observations. The new $Q_{\phi}(H)$ observations correspond to a 3-year time difference follow-up. These new data, along with a re-analysis of the ALMA 230\,GHz continuum, allow us to reach the following conclusions:

\begin{itemize}
 \item We found intensity dips in the resolved ALMA imaging of the radio continuum, confirming the predictions from paper\,I. The locations of the northern radio decrements in both 230\,GHz and 350\,GHz are fairly aligned, with position angles of 316.3\,deg and 316.9\,deg, east of north, respectively. The southern radio decrements, which seem broader, are both aligned at a position angle of 152.9\,deg east of north. 
\item Provided a retrograde disc (rotating clockwise), the temperature decrements are leading in the direction of rotation relative to the $Q_{\phi}(H)$ decrements, by $\eta$\,=\,14.95\,deg in the location of the northern dips and $\eta$\,=\,7.92\,deg in the location of the southern dips.
\item Geometrical models of $Q_{\phi}(H)$ provide support for a misaligned inner disc, with a relative inclination of $\xi$\,=\,21.4\,$^{+6.7}_{-8.3}$\,deg. The position angle joining both $Q_{\phi}(H)$ decrements results in a shift along the direction west of north, compared to previous observations presented in paper\,I. This is reflected in the morphological change in the location of the southern decrement, which also implies a variation in the orientation of the inner disc between the two $Q_{\phi}(H)$ epochs.
\item The intensity dips may correspond to temperature decrements, since the spectral index map between 230\,GHz and 350\,GHz shows values $\alpha \sim$\,2 in the location of the shadows, which is consistent with optically thick emission.
\item Disc gas mass estimates using the angular shifts between the radio and IR decrements, and for a standard dust population, give surface density values of $\Sigma_g$\,=\,117\,$\pm$\,10 g/cm$^2$ for the northern decrement, and $\Sigma_g$\,=\,48\,$\pm$\,10 g/cm$^2$ for the southern decrement. A lower value of $\Sigma_g$ is consistent with a broader and smoother temperature decrement in the south.
\item The spectral index values $\alpha \sim$\,3.3 in the inner cavity, where we observe a faint mm-wavelength dust pedestal at the edge, are typical for the ISM and suggest optically thin emission due to the presence of smaller grains compared to the outer regions of the disc where $\alpha \sim$\,2. 
\item The outer continuum halo shows spectral index values of $\alpha \sim$\,2, which suggests the presence of fluffy large grains. 
\item The observed mm-wavelength outer dust halo shows spectral index values around $\alpha \sim$\,2, which can be interpreted as optically thick emission or as a shallow dust opacity index. Multi-wavelength data of DoAr\,44 is required to further constrain the spectral index variations in the disc.

\end{itemize}

\section*{Acknowledgments} 
We thank Prof. Dr. Hans Martin Schmid for constructive comments on the SPHERE observations. We thank the anonymous referee for helpful comments. The authors acknowledge financial support from ANID – Millennium Science Initiative Program – Center Code NCN2021\_080. C.A-T acknowledges support from the Millennium Nucleus on Young Exoplanets and their Moons (YEMS). S.C. acknowledges support from Agencia Nacional de Investigaci\'on y Desarrollo de Chile
(ANID) given by FONDECYT Regular grant 1211496, and ANID project Data Observatory Foundation DO210001. S.P. acknowledges support from FONDECYT Regular grant 1191934. P.W. acknowledges support from FONDECYT grant 3220399. L.C. acknowledges financial support from FONDECYT grant $\#$1211656. A.Z. acknowledges support from the FONDECYT Iniciaci\'on en investigaci\'on project number 11190837. SM is supported by a Royal Society University Research Fellowship (URF-R1-221669). This work used the Strelka cluster (FONDEQUIP project EQM140101) hosted at DAS/U. de Chile. This paper makes use of the following ALMA data: ADS/JAO.ALMA $\#$2018.1.00028.S. ALMA is a partnership of ESO (representing its member states), NSF (USA) and NINS (Japan), together with NRC (Canada), MOST and ASIAA (Taiwan), and KASI (Republic of Korea), in cooperation with the Republic of Chile. The Joint ALMA Observatory is operated by ESO, AUI/NRAO and NAOJ. The National Radio Astronomy Observatory is a facility of the National Science Foundation operated under cooperative agreement by Associated Universities, Inc.

\section*{Data Availability} 
The reduced ALMA and SPHERE data presented in this article are available upon reasonable request to the corresponding author. The original or else non-standard software packages underlying the analysis are available at the following URLs: MPolarMaps (\url{https://github.com/simoncasassus/MPolarMaps}, \citealt{casassus21}), uvmem (\url{https://github.com/miguelcarcamov/gpuvmem}, \citealt{carcamo18}), VisAlign (\url{https://github.com/simoncasassus/VisAlign}) and SNOW (\url{https://github.com/miguelcarcamov/snow}).

\label{lastpage}

\bibliographystyle{mn2e}

\bibliography{doar44}
\bsp

\end{document}